# Critical Confinement and Elastic Instability in Thin Solid Films

**Animangsu Ghatak[1,*] and Manoj K. Chaudhury[2]**

[1]Department of Chemical Engineering
Indian Institute of Technology, Kanpur 208016, India,
[2]Department of Chemical Engineering, Lehigh University,
Iacocca Hall, 111 Research Drive, Bethlehem, Pennsylvania 18015-4791

**Abstract**: When a flexible plate is peeled off a thin and soft elastic film bonded to a rigid support, uniformly spaced fingering patterns develop along their line of contact. While, the wavelength of these patterns depends only on the thickness of the film, their amplitude varies with all material and geometric properties of the film and that of the adhering plate. Here we have analyzed this instability by the regular perturbation technique to obtain the excess deformations of the film over and above the base quantities. Furthermore, by calculating the excess energy of the system we have shown that these excess deformations, associated with the instability, occur for films which are critically confined. We have presented two different experiments for controlling the degree of confinement: by pre-stretching the film and by adjusting the contact width between the film and the plate.

## Introduction:

Pattern formation by self organization is a subject of much interest because of its immense scientific and technological importance. While examples of instability driven evolution of such patterns are abound in dynamic systems involving viscous and visco-elastic materials [1-10], such instances reported for purely elastic solids are rather scanty. Recently, such a pattern forming system has been identified [11-13] with thin soft elastic films confined between rigid or flexible substrates. Here the patterns of instability appear when a flexible plate is peeled off a layer of elastic adhesive bonded to rigid substrate, resulting in undulations along their line of contact. Neither the appearance of these patterns nor their wavelength depends on the rate of peeling thus remaining independent of the dynamics of the system. Indeed, the morphology does not exhibit any temporal evolution even when the contact line comes to a complete rest. The elastic nature of the film allows us to form the patterns repeatedly on the same film, so that similar patterns can be replicated many times. Looking at the prospect of this instability being used as a powerful pattern forming tool, we have studied it extensively [14] in variety of experimental geometries as well as with adhesives and adherents with varying material and physical properties. We have developed also methods to fix permanently these patterns [15].

In these experiments, a layer of elastic film of thickness $h$ and shear modulus $\mu$ remains strongly bonded to a rigid substrate and a microscope cover slip of flexural rigidity $D$ is peeled off it by inserting a spacer at the opening of the crack. The patterns appear in the form of well defined undulation at the contact line. While the wavelength $\lambda$ of these waves increases linearly with the thickness $h$ of the film, remaining independent of its shear modulus $\mu$ and the flexural rigidity $D$ of the plate, amplitude $A$ varies rather nonlinearly with these parameters. For the sake of systematic presentation of these results, we have introduced a confinement parameter $\varepsilon = hq$ defined as [16] the ratio of two different length-scales: thickness $h$ and $q^{-1} = \left(Dh^3 / 3\mu\right)^{1/6}$ [17,18], the latter being the stress decay length along the film/plate interface from the contact line (along negative $x$ direction in fig. 1a). These definitions imply that lower is the value of $\varepsilon$, longer is the stress decay length for a film of a given thickness; hence more confined is the film. For large values of $\varepsilon$, i.e. low levels of confinement, the film can compensate its stretching perpendicular to the interface via lateral Poisson contraction. However, when $\varepsilon$ decreases below a critical value ($\varepsilon < 0.35$), the films cannot afford a large scale Poisson contraction. Then, in order to accommodate lateral contraction at a local level, the contact line turns undulatory to engender uniformly spaced fingers and cavities. In this report we present two different experimental schemes to demonstrate how the confinement can be controlled in a systematic way. In one experiment, a flexible plate is lifted off a thin elastic film from both of its ends thus effecting an adjustable contact width between the two, while in the other, it is lifted off a film which is pre-stretched uniaxially. The confinement of the film increases in both of these experiments, with increase in the contact width and the extent of stretching.

**Experiment with double spacers:**

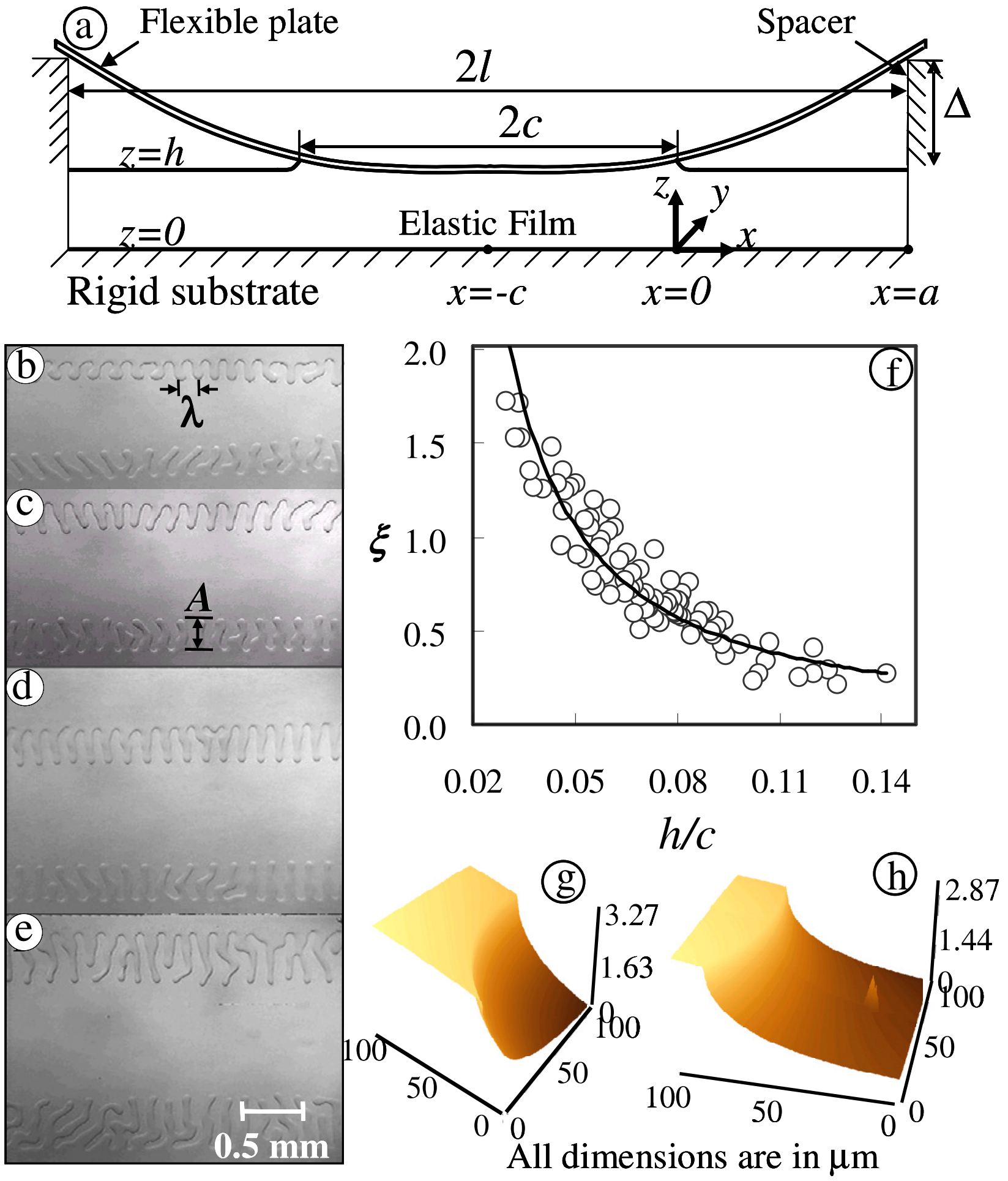

**Fig 1. (a) A 2d sketch (figure not drawn according to scale) of the experiment in which a flexible glass plate (silanized with self assembled monolayers of hexadecyltrichlorosilane) is peeled off an elastic film of crosslinked poly(dimethylsiloxane) (PDMS) with two spacers. (b)-(e) are the typical images of the contact region as the distance between the spacers ($2l$) is progressively increased: $2l = 14.5, 19.5, 22.5$ and $27.6$ mm respectively. These micrographs are obtained with a film of $\mu = 0.2$ MPa, $h = 40\,\mu$m and a flexible plate $D = 0.02$ Nm. (f) Amplitude data from experiments with films of shear modulus $\mu = 0.2 - 1.0$ MPa and thickness $h = 40 - 200$ µm and flexible plates of rigidity: $D = 0.02 - 0.06$ Nm are scaled as $\xi = Aq$ and plotted against the quantity $h/c$. The solid line is a guide to the eye. (g)-(h) These instability patterns can be generated on a partially crosslinked PDMS film, which can then be fully crosslinked in order to permanently fix these patterns. Typical atomic force microscopy images of such a permanently fixed pattern**

**of the film ($\approx 150$ μm) in close proximity to the fingers suggest that maximum normal strain of the film is $< 0.02$. Figure (g) is the image of the region adjoining two fingers, whereas figure (h) is that of the tip of finger.**

**Materials.** We obtained the glass slides (Corning microslides) and cover slips (Corning cover plates) from Fisher Scientific, USA. We cleaned the glass slides in a Harrick plasma cleaner (model PDC-23G, 100W) before surface treatment. The material for film preperation, i.e., vinyl-endcapped poly(dimethylsiloxane) oligomers of different chain lengths, platinum catalyst, and the methylhydrogen siloxane cross-linker, were obtained as gifts from Dow Corning Corp., Midland, MI, USA. We used also two sets of filler gauges of various thicknesses which were purchased from a local auto-parts shop. The instability patterns were observed with a Nicon Diaphot inverted microscope equipped with a CCD camera and a video recorder.

**Method**

Figure 1(a) depicts the schematic of the first experiment in which the flexible plate is detached from the bonded elastic layer, by inserting two spacers of height $\Delta$ on two sides of the plate/film interface. The distance between the spacers is $2l$. The spacers generate two propagating cracks at the interface and peel the flexible plate off the adhesive film from both its ends. A finite contact width $2c$ is attained following the equilibrium of forces which include the adhesion and elastic forces in the adhesive and the adherent. When the spacers remain far apart, so that $2l$ and $2c$ both tend to infinity, the experiment represents the limiting case in which a single spacer is used to lift the flexible plate off the film from one of its ends [11]. However, as the distance $2l$ is decreased, the contact width $2c$ shrinks thereby increasing the curvature of the plate. Finally a distance is reached at which the stress required to bend the plate exceeds the adhesion strength of the interface and the plate no longer obliges to remain stuck to the film. Here we will present a systematic analysis of this experiment in order to rationalize our experimental observations.

**Preamble:**

Since no dynamics is involved in the formation of these unconditionally stable patterns, we develop our arguments on the premise that their final states are attained by the minimization of the overall energy of the system. In particular, we consider four types of energy: the bending energy of the contacting plate, the elastic energy of the film, the adhesion energy at the interface of the plate and the film and the surface energy associated with the creation of curved surface near the finger region. This last contribution is however negligible in comparison to the elastic energy of the film as the two scales with the ratio [11]: $\gamma/\mu h \approx 10^{-5}$ where, $\gamma$ is the surface energy of the film and $\mu$ is its shear modulus. Furthermore, following observations from experiments, the contacting plate has been assumed to bend only in the direction of propagation of the contact line (i.e. along $x$) remaining uniform along the wave-vector (i.e. along $y$); its contribution to the total energy has been accounted for accordingly. While, due to the peeling action on the plate, the film deforms normal to the surface, due to its own incompressibility, it sags in the vicinity of the contact line. For a thick film, the shear deformation occurs in both the $xy$ and $yz$ planes i.e. in planes normal to the $z$ and the $x$ axes respectively. The latter deformation causes sagging in region ahead of the line of contact of the film and the plate, which is enough for compensating the normal deformation of the film. This kind of deformation does not cause undulation of the contact line. However, when the film is thin, the hydrostatic stress in it causes shear deformations also in the $xz$ (i.e. normal to $y$ axes) plane. It is this additional lateral deformation that results in undulation of the contact line, which is accounted for in the energy calculations. The remaining component of the energy is the work of adhesion, which depends on the area of contact between the film and the cover plate. Previous studies [13] indicated that the magnitude of the adhesion energy does not affect the wavelength of instability. These linear analyses however provide no information about the amplitude of instability. In the current problem, as the overall energy is minimized with respect to wavelength and amplitudes as variables, a weak non-linearity is naturally invoked that presupposes a geometric and energetic relationship between the above two variables. This non-linearity is valuable in generating the bifurcation diagram of the morphology of fingers in which the total energy goes through a minimum at specific amplitude and wavelength for a given confinement parameter. Our analysis shows that such minima exist only when the film is sufficiently confined.

**Governing Equations and Boundary Conditions:**

In the absence of any body force, the stress and the displacement profiles in the elastic adhesive are obtained by solving the stress equilibrium and the incompressibility relations for an incompressible elastic material:

$$-\overline{\nabla} p + \mu \nabla^2 \overline{u} = 0 \text{ and } \overline{\nabla} \cdot \overline{u} = 0 \tag{1}$$

where $p$ is the pressure field and $\overline{u} = u\overline{e_x} + v\overline{e_y} + w\overline{e_z}$ is the displacement vector in the film with $x$, $y$ and $z$ being respectively the direction of propagation of the contact line, direction of the wave vector and the thickness co-ordinate of the film. To be precise in the $x = 0$ line, i.e. the $y$ axis goes through the tip of the wavy contact line, whose position is represented by $\eta(y)$. Equations 1 are solved with the following boundary conditions [11-15]:

(a) The film remains perfectly bonded to the rigid substrate so that at $z = 0$, displacements $\bar{u} = 0$.

(b) The flexible contacting plates are coated with self assembled monolayer (SAM) of hexadecyltrichlorosilane molecules which allow for partial slippage at the interface of the film and the flexible plate i.e. at $z = h + \psi$, where, $\psi$ is the vertical displacement of the interface measured from the un-deformed surface of the film. While similar surface treatments can alter the interfacial friction as evident in our earlier experiments [14, 19], the wavelength of perturbations does not depend on the level of frictional resistance at the surface. Furthermore, our calculation in reference 18 shows that the two extreme conditions of perfect bonding and infinite slippage both lead to similar values for work of adhesion at the interface. Hence, for the sake of simplicity, we assume frictionless contact at the film-flexible plate interface so that the shear stress $\sigma_{xz}\big|_{z=h+\psi} = \sigma_{yz}\big|_{z=h+\psi} = 0$.

(c) At the film-flexible plate interface ($x < \eta(y)$), the traction on the film is equal to the bending stress on the plate which in our experiments bends through a very small angle $< 1^{\circ}$. Hence under small bending approximations, we have $\sigma_{zz}\big|_{x<\eta(y),z=h+\psi} + D\Delta^2\psi = 0$ i.e.

$\left(-p+2\mu\frac{\partial w}{\partial z}\right)\Big|_{x<\eta(y),z=h+\psi}+D\Delta^2\psi=0$. At $\eta(y)<x<a$, the film and the plate are free of any traction, so that $\sigma_{zz}\big|_{\eta(y)<x<a,z=h+\psi}=0$ and $D\Delta^2\psi=0$. Here $a$ represents the distance of the spacer from $x=0$. These boundary conditions applied at the two different regions then should lead to discontinuity of the normal stress across the contact line. However this discontinuity does not occur, because within a very small distance away from the contact line $x>\eta(y)+$, the film and the plate remains subjected to distance dependent intermolecular forces of adhesion which drops continuously to zero leading to a continuous variation of the normal stress. This point is further elaborated in the following paragraph.

(d) While there is no ambiguity regarding the above set of boundary conditions applied in the two regions, i.e. at $x<\eta(y)$ and at $\eta(y)<x<a$, the nature of the boundary condition at $x=\eta(y)$ is less clear and has been a subject of considerable discussions in recent years [20-22]. Recently Mokhter and Mahadevan [23] treated the crack tip instability problem by considering a single plate in contact with a thin soft elastic film. At the boundary line, they considered the condition of the classical singularity of normal stress, which led to the discontinuity of the displacement derivatives of the plate and the film. However, Maugis [20], who discussed this issue rather extensively, came to the conclusion that singularity of the stress field as developed by the remote loading at the crack tip is cancelled by another singularity due to internal loading resulting from the cohesive stresses at the crack tip. The cancellation of the two singularities leads to a smooth variation of all the slopes at the crack tip region. The problem has also been addressed recently [20, 22] from the point of view that the stress at the crack tip for soft polymer films cannot exceed the value of its elastic modulus because of the fact that the polymer chains bridging the two surfaces undergo thermal fluctuations. While the reference 21 treats the problem using rigorous statistical mechanics, here we address the problem in a somewhat simplified way by considering that any one of the bridging chains can be either in the attached or in a detached state. At equilibrium, the areal density of the bonded chains is given by

$$\Sigma_b = \frac{\Sigma_0}{1 + \exp\left( \frac{k_s \delta^2}{2k_B T} - \frac{\varepsilon_A}{k_B T} \right)} \tag{2}$$

where, $\Sigma_0$ is the total areal density of bonded and unbonded chains, $k_s$ is the spring constant of the chain, $\varepsilon_A$ is the energy of adsorption per chain and $\delta$ is the extension of the chain. The stress is obtained by multiplying $\Sigma_b$ with the spring force ($k_s \delta$). Furthermore, recognizing that the spring constant ($k_s$) of a Gaussian chain is given by $k_B T / n_s l_s^2$ ($l_s$ is the statistical segment length and $n_s$ is the numbers of statistical segments per chain) and the elastic modulus of the network $E \approx k_B T / n_s l_s^3$, the normal stress at the interface can be written as:

$$\sigma = \frac{E \sqrt{\frac{2\phi}{k_B T}}}{1 + \exp\left( \frac{\phi - \varepsilon_A}{k_B T} \right)} \tag{3}$$

where, $\phi = k_s \delta^2 / 2$. The above representation suggests that the normal stress $\sigma$ converges asymptotically to zero for $\phi$ approaching either zero or infinity and attains a maximum at an intermediate value of $\phi$ . Far away from the crack tip i.e. at $x \to -\infty$, at the interface of the film and the plate no polymer chain is stretched so that the normal stress is zero. Within the cohesive zone of the crack tip, as we traverse in the other direction, the polymer chains get more and more stretched, thus $\phi$ increases. However, this increase in $\phi$ is accompanied by the decrease in number of the bonded chains resulting in the decrease of the normal stress as dictated by equation 2. Within these two zones, in the open mouth of the crack, very close to the crack tip i.e. at $x \to \eta(y)-$ the normal stress $\sigma$ goes through a maximum. However, this stress can not exceed the maximum cohesive stress at the contact line, which is only a fraction of the elastic modulus $E$ as predicted by equation 3 for representative values of the adsorption energy $\varepsilon_A$ for van der Waals interactions. For these soft materials, therefore, the normal stress rises smoothly to a finite maximum value at the crack tip and then it falls smoothly within a very small distance of the contact line. Within the context of the continuum mechanics formalism, we use the boundary condition that the hydrostatic tension is maximum in the near vicinity of the contact line, i.e. $\left.\frac{\partial p}{\partial x}\right|_{x=\eta(y)} = 0$.

Using the above boundary conditions we proceed to solve equations 1. We first write the displacements and the co-ordinates in dimensionless form using the following two length scales: $q^{-1} = \left(Dh^3/3\mu\right)^{1/6}$ as the characteristic length along $x$ obtained naturally from the analysis presented in references 14 and 15 and the thickness $h$ of the film as that along $y$ following observation that the wavelength of instability $\lambda$ varies linearly with the thickness of the film [11]. Furthermore, thickness $h$ is also the characteristic length along the $z$ axis. Using these characteristic lengths we obtain the following dimensionless quantities:

$X = xq,\ Y = y/h,\ Z = z/h,\ U = uq,\ V = v/h,\ W = w/h$

We write the stresses too in dimensionless form by dividing them with $\mu/\varepsilon^2$ as the characteristic pressure:

$P = p/\left(\mu/\varepsilon^2\right),\ \Sigma_{XZ} = \sigma_{xz}/\left(\mu/\varepsilon^2\right),\ \Sigma_{YZ} = \sigma_{yz}/\left(\mu/\varepsilon^2\right).$

Using these dimensionless quantities, the stress equilibrium and incompressibility relations are written as:

$$\frac{\partial P}{\partial X} = \varepsilon^2 \frac{\partial^2 U}{\partial X^2} + \frac{\partial^2 U}{\partial Y^2} + \frac{\partial^2 U}{\partial Z^2}, \quad \frac{\partial P}{\partial Y} = \varepsilon^4 \frac{\partial^2 V}{\partial X^2} + \varepsilon^2 \left( \frac{\partial^2 V}{\partial Y^2} + \frac{\partial^2 V}{\partial Z^2} \right),$$

$$\frac{\partial P}{\partial Z} = \varepsilon^4 \frac{\partial^2 W}{\partial X^2} + \varepsilon^2 \left( \frac{\partial^2 W}{\partial Y^2} + \frac{\partial^2 W}{\partial Z^2} \right)$$

and $\dfrac{\partial U}{\partial X} + \dfrac{\partial V}{\partial Y} + \dfrac{\partial W}{\partial Z} = 0$ (4)

And the boundary conditions result in

(a) $U(Z=0) = V(Z=0) = 0$

(b) $\Sigma_{XZ}(X,Y,Z=1+\Psi) = 0 = \Sigma_{YZ}(X,Y,Z=1+\Psi)$

or, $\left( \dfrac{\partial U}{\partial Z} + \varepsilon^2 \dfrac{\partial W}{\partial X} \right) = \left( \dfrac{\partial V}{\partial Z} + \dfrac{\partial W}{\partial Y} \right) = 0$

(c) $P(Z=1+\Psi) = 2\varepsilon^2 \left.\dfrac{\partial W}{\partial Z}\right|_{Z=1+\Psi} + 3\left( \dfrac{\partial^4 \Psi}{\partial X^4} + \dfrac{2}{\varepsilon^2}\dfrac{\partial^4 \Psi}{\partial X^2 \partial Y^2} + \dfrac{1}{\varepsilon^4}\dfrac{\partial^4 \Psi}{\partial Y^4} \right)$ at $X < \mathrm{N}(Y)$,

(d) $0 = \dfrac{\partial^4 \Psi}{\partial X^4}$ at $\mathrm{N}(Y) < X < aq$

(e) $\left.\dfrac{\partial P}{\partial X}\right|_{X=\mathrm{N}(Y)} = 0$ (5)

where $aq$ is the dimensionless distance of the spacer from $X = 0$, $\Psi$ is the dimensionless vertical displacement of plate and $X = \mathrm{N}(Y)$ represents the position of the contact line in the dimensionless form. We assume that the solutions of displacements and pressure consist of two components: the base solutions which vary only along the $X$ and $Z$ axes and remain uniform along $Y$, and a perturbation term which appear over and above these base solutions and incorporate the spatial variation of the displacements along the $Y$ axis. The general form of these solutions are expressed as [24] $T = T_0(X,Z) + \varepsilon^2 T_1(X,Y,Z) + \varepsilon^4 T_2(X,Y,Z) + \ldots$ where $T = P, U, V$ and $W$. Similarly the vertical displacement $\Psi$ of the plate is expanded as $\Psi = \Psi_0(X) + \varepsilon^2 \Psi_1(X,Y) + \varepsilon^4 \Psi_2(X,Y) + \ldots$ Here, the base solutions are of order $\varepsilon^0$ and the perturbed ones are of order $\varepsilon^2$, $\varepsilon^4$, etc. At this juncture, it is worthwhile to point out the physical nature of the perturbation as $\varepsilon^2$ approaches zero which corresponds to the stress decay length $q^{-1}$ approaching infinity. For any finite deformation of the elastomeric film at the crack tip region, the above condition implies that the local radius of curvature of the cantilever plate also approaches infinity or that the slope of the cantilever plate is vanishingly small. As we show below, this base state solution follows from the premise that the pressure and the displacements remain uniform along the $y$ axis and the classical lubrication approximation is applicable. Later, we seek a solution of equation 4 (or equation 6, below) by considering periodic perturbations of the field variables along the $y$ direction. This leads to a geometric perturbation of the contact line, where the base state corresponds to a vanishing wave vector of the periodic perturbation (i.e. straight contact line).

Notice that in a typical experiment $h = 50\,\mu\text{m}$, $\mu = 1.0\,\text{MPa}$ and $D = 0.02\,\text{Nm}$, so that $\varepsilon^2 = 0.03 << 1$, is small enough to allow perturbation expansion for the displacement and pressure fields as powers of $\varepsilon^2$. When we insert these definitions in equation 4 and separate the base ($Y$ independent) and the perturbation ($Y$ dependent) terms, we obtain

$$\begin{aligned} -\frac{\partial P_0}{\partial X} + \varepsilon^2 \frac{\partial^2 U_0}{\partial X^2} + \frac{\partial^2 U_0}{\partial Z^2} = \varepsilon^4 \frac{\partial^2 V_0}{\partial X^2} + \varepsilon^2 \frac{\partial^2 V_0}{\partial Z^2} = 0 \\ -\frac{\partial P_0}{\partial Z} + \varepsilon^4 \frac{\partial^2 W_0}{\partial X^2} + \varepsilon^2 \frac{\partial^2 W_0}{\partial Z^2} = \frac{\partial U_0}{\partial X} + \frac{\partial W_0}{\partial Z} = 0 \end{aligned} \tag{6}$$

and

$$\begin{aligned} \varepsilon^2 \frac{\partial P_1}{\partial X} + \varepsilon^4 \frac{\partial P_2}{\partial X} &= \varepsilon^2\left(\frac{\partial^2 U_1}{\partial Y^2} + \frac{\partial^2 U_1}{\partial Z^2}\right) + \varepsilon^4\left(\frac{\partial^2 U_1}{\partial X^2} + \frac{\partial^2 U_2}{\partial Y^2} + \frac{\partial^2 U_2}{\partial Z^2}\right) + \varepsilon^6 \frac{\partial^2 U_2}{\partial X^2} \\ \varepsilon^2 \frac{\partial P_1}{\partial Y} + \varepsilon^4 \frac{\partial P_2}{\partial Y} &= \varepsilon^4\left(\frac{\partial^2 V_1}{\partial Y^2} + \frac{\partial^2 V_1}{\partial Z^2}\right) + \varepsilon^6\left(\frac{\partial^2 V_1}{\partial X^2} + \frac{\partial^2 V_2}{\partial Y^2} + \frac{\partial^2 V_2}{\partial Z^2}\right) \\ \varepsilon^2 \frac{\partial P_1}{\partial Z} + \varepsilon^4 \frac{\partial P_2}{\partial Z} &= \varepsilon^4\left(\frac{\partial^2 W_1}{\partial Y^2} + \frac{\partial^2 W_1}{\partial Z^2}\right) + \varepsilon^6\left(\frac{\partial^2 W_1}{\partial X^2} + \frac{\partial^2 W_2}{\partial Y^2} + \frac{\partial^2 W_2}{\partial Z^2}\right) \\ 0 &= \frac{\partial U_1}{\partial X} + \frac{\partial V_1}{\partial Y} + \frac{\partial W_1}{\partial Z} \end{aligned} \tag{7}$$

Which are solved using the following b.c. derived from equations 5a-e:

(a) at $Z = 0$, $U_0 = W_0 = 0$

and $\varepsilon^2 U_1 + \varepsilon^4 U_2 = \varepsilon^2 V_1 + \varepsilon^4 V_2 = \varepsilon^2 W_1 + \varepsilon^4 W_2 = 0$

(b) at $Z = 1 + \Psi$, $\frac{\partial U_0}{\partial Z} + \varepsilon^2 \frac{\partial W_0}{\partial X} = \frac{\partial W_0}{\partial Y} = 0$

and $\varepsilon^2 \frac{\partial U_1}{\partial Z} + \varepsilon^4\left(\frac{\partial U_2}{\partial Z} + \frac{\partial W_1}{\partial X}\right) = \varepsilon^2\left(\frac{\partial V_1}{\partial Z} + \frac{\partial W_1}{\partial Y}\right) + \varepsilon^4\left(\frac{\partial V_2}{\partial Z} + \frac{\partial W_2}{\partial Y}\right) = 0$

c) at $Z = 1 + \Psi$ $P_0 = 3\frac{\partial^4 \Psi_0}{\partial X^4} + 2\varepsilon^2 \frac{\partial W_0}{\partial Z}$, $X < 0$

$$\varepsilon^2 P_1 + \varepsilon^4 P_2 = \frac{1}{\varepsilon^2}\frac{\partial^4 \Psi_1}{\partial Y^4} + \left(2\frac{\partial^4 \Psi_1}{\partial X^2 \partial Y^2} + \frac{\partial^4 \Psi_2}{\partial Y^4}\right)$$

and $\qquad X < \mathrm{N}(Y)$

$$+\varepsilon^2\left(3\frac{\partial^4 \Psi_1}{\partial X^4} + 2\frac{\partial^4 \Psi_2}{\partial X^2 \partial Y^2}\right) + \varepsilon^4\left(3\frac{\partial^4 \Psi_2}{\partial X^4} + 2\frac{\partial W_1}{\partial Z}\right)$$

(d) at $Z = 1 + \Psi$ $\qquad 0 = \dfrac{\partial^4 \Psi_0}{\partial X^4} \qquad 0 < X < aq$

and $\qquad 0 = \varepsilon^2 \dfrac{\partial^4 \Psi_1}{\partial X^4} + \varepsilon^4 \dfrac{\partial^4 \Psi_2}{\partial X^4} \qquad \mathrm{N}(Y) < X < aq$

(e) $0 = \left.\dfrac{\partial P_0}{\partial X}\right|_{X=0}$ and $0 = \varepsilon^2 \left.\dfrac{\partial P_1}{\partial X}\right|_{X=\mathrm{N}(Y)} + \varepsilon^4 \left.\dfrac{\partial P_2}{\partial X}\right|_{X=\mathrm{N}(Y)}$ (8)

**Solution of Equation 6 using Long Scale Approximation:**

We solve equation 6 in the light of boundary conditions in equation 8. However, the characteristic length-scales along $X$ and $Z$ axes are so far apart that $\varepsilon^2 << 1$ as noted earlier. This allows us to simplify these equations using the long-scale approximations [21]: $-\dfrac{\partial P_0}{\partial X} + \dfrac{\partial^2 U_0}{\partial Z^2} = -\dfrac{\partial P_0}{\partial Z} = \dfrac{\partial U_0}{\partial X} + \dfrac{\partial W_0}{\partial Z} = 0$; at $Z = 0$, $U_0 = W_0 = 0$ and at $Z = 1 + \Psi$, $\dfrac{\partial U_0}{\partial Z} = \dfrac{\partial W_0}{\partial Y} = 0$. When these equations are integrated over $0 < Z < 1 + \Psi_0$, following relations for the displacements in the film results,

$$U_0 = \frac{\partial P_0}{\partial X}\left(\frac{Z^2}{2} - (1 + \Psi_0)Z\right)$$
$$W_0 = -\frac{\partial^2 P_0}{\partial X^2}\frac{Z^3}{6} + \frac{Z^2}{2}\left(\frac{\partial^2 P_0}{\partial X^2}(1 + \Psi_0) + \frac{\partial P_0}{\partial X}\Psi_0\right) \qquad (9)$$

Hence, the vertical displacement $\Psi_0 = W_0|_{Z=1+\Psi_0}$ of the interface is obtained as

$$
\begin{aligned}
\Psi_0 &= \frac{\partial^2 P_0}{\partial X^2}\frac{(1+\Psi_0)^3}{3} + \frac{\partial P_0}{\partial X}\frac{\Psi_0(1+\Psi_0)^2}{2} \\
&= \frac{\partial^6 \Psi_0}{\partial X^6}(1+\Psi_0)^3 + \frac{\partial^5 \Psi_0}{\partial X^5}\frac{3\Psi_0(1+\Psi_0)^2}{2}
\end{aligned}
\tag{10}
$$

While the nonlinearity of the above equation renders it non-amenable for analytical solutions, we will simplify it by noting that in all our experiments the dimensionless vertical displacement $\Psi_0 < 0.02$. Thus linearization of the above equation results in

$$
\Psi_0 - \frac{\partial^6 \Psi_0}{\partial X^6} = 0 \text{ at } X < 0 \tag{11}
$$

At $0 < X < aq$ there is no traction on plate i.e. $\frac{\partial^4 \Psi_0}{\partial X^4} = 0$. Equation 11 is solved using the boundary condition that about the centerline $X = -cq$ of the contact area, displacement $\Psi_0$ and the bending moment and normal stress on plate are continuous, so that $\frac{\partial \Psi_0}{\partial X} = \frac{\partial^3 \Psi_0}{\partial X^3} = \frac{\partial^5 \Psi_0}{\partial X^5} = 0$ and that it is freely supported at $X = aq$ by the spacer ($\Psi_0|_{X=aq} = \overline{\Delta} = \frac{\Delta}{h}$, $\frac{\partial^2 \Psi_0}{\partial X^2} = 0$). Finally at the vicinity of the contact line ($X = 0$) displacement, slope, bending moment and vertical shear force are continuous, so that $\Psi_0|_{X=0-} = \Psi_0|_{X=0+}$, $\left.\frac{\partial \Psi_0}{\partial X}\right|_{X=0-} = \left.\frac{\partial \Psi_0}{\partial X}\right|_{X=0+}$, $\left.\frac{\partial^2 \Psi_0}{\partial X^2}\right|_{X=0-} = \left.\frac{\partial^2 \Psi_0}{\partial X^2}\right|_{X=0+}$ and $\left.\frac{\partial^3 \Psi_0}{\partial X^3}\right|_{X=0-} = \left.\frac{\partial^3 \Psi_0}{\partial X^3}\right|_{X=0+}$.

Maximal tensile stress at the contact line results in $\left.\frac{\partial P_0}{\partial X}\right|_{X=0} = \left.\frac{\partial^5 \Psi_0}{\partial X^5}\right|_{X=0} = 0$.

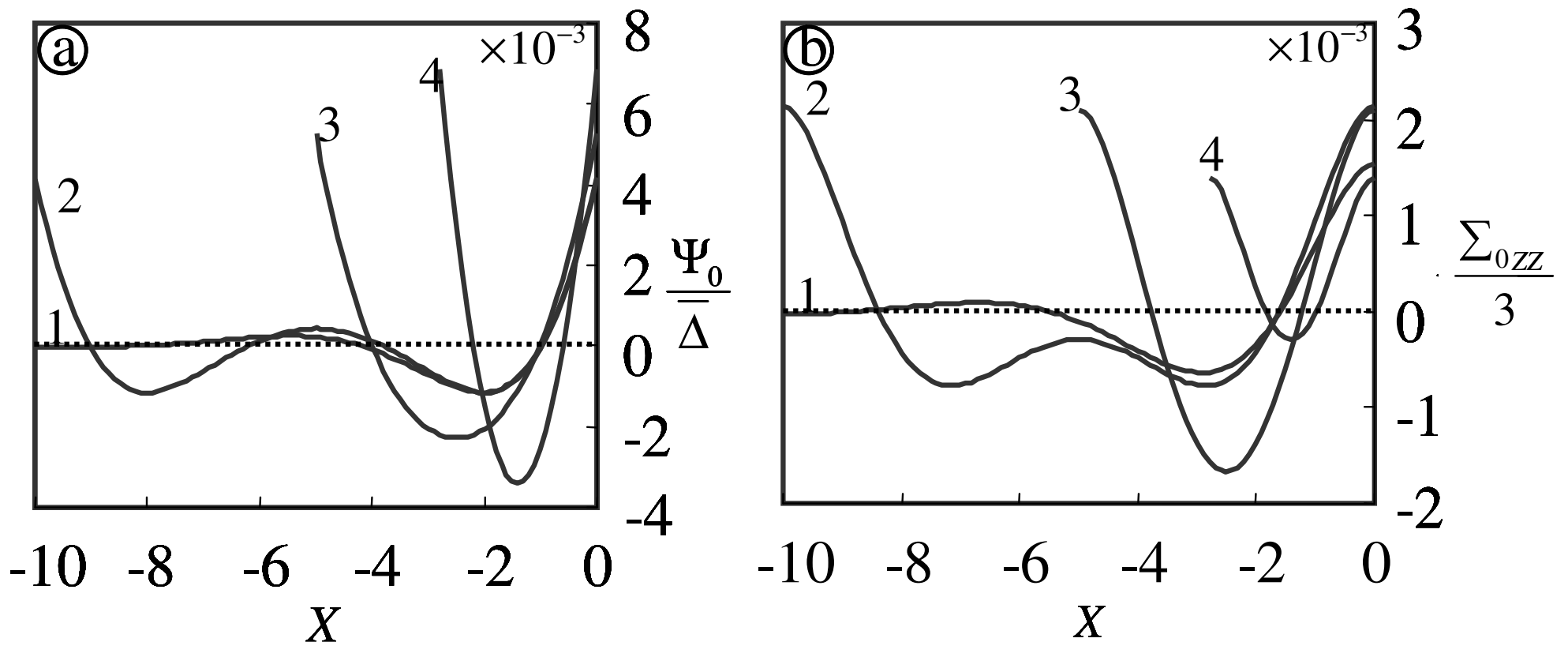

**Fig 2. The plots depict the dimensionless vertical displacement $\Psi_0 / \overline{\Delta}$ and normal stress $\Sigma_{0ZZ} / 3$ in the elastic (PDMS) film in experiments of figure 1(a). The profile 1 obtained with inter-spacers distance $2l \rightarrow \infty$ represents the limiting case in which single spacer peels off the plate. Curves 2-4, corresponding to ($aq = 25, cq = 5$), ($15, 2.5$) and ($15, 1.4$) are obtained when $2l$ is progressively reduced thereby decreasing the width of the contact area.**

In figure 2 we plot the numerical solutions of dimensionless vertical displacement $\Psi_0$ and normal stress $\Sigma_{0ZZ} = \varepsilon^2 \sigma_{0_{zz}} / \mu$ ($\sigma_{0_{zz}}$: base components of the normal stress on film) with respect to dimensionless distance $X$ for representative set of values for dimensionless distances $aq = 15 - 25$ and $cq = 2.5 - 10$. Curve 1 represents the limiting case of semi-infinite contact area, for which the displacement and stress profiles remain oscillatory with exponentially diminishing amplitude away from the contact line [18]; whereas curve 2 represents an intermediate situation in which contact area decreases as the spacers are brought closer. Here too we see oscillatory profiles with diminishing amplitude, but number of oscillation decreases. Finally for curves 3 and 4, the contact width is so small that a region of positive $\Psi_0$ and $\Sigma_{0ZZ}$ close to $X = 0$ is followed by a region of negative values of $\Psi_0$ and $\Sigma_{0ZZ}$. Using the expressions of these displacement profiles, we can estimate the total energy of the system which consists of the elastic energy of the film, the bending energy of the plate and the interfacial work of adhesion $W_A$. The dimensionless form of the total energy is obtained as:

$$\Pi_0 = \frac{\varepsilon^3}{2}\left( \int_{-cq}^{aq} \left( \frac{d^2\Psi_0}{dX^2} \right)^2 dX + \frac{1}{6} \int_{-cq}^{aq} \int_0^1 \left( \frac{\partial U_0}{\partial Z} + \varepsilon^2 \frac{\partial W_0}{\partial X} \right)^2 dXdZ \right) + \frac{W_A h^2}{D} \frac{aq}{\varepsilon} \qquad (12)$$

which is a function of the four dimensionless quantities $aq, \varepsilon, \frac{W_A h^2}{D}$ and $cq$, the numerical value of which are such that the total energy of the system is minimum. This later condition allows us to obtain one of the parameters for a given set of three other parameters. For example, for a given set of values of $cq, \varepsilon$ and $\frac{W_A h^2}{D}$ we obtain $aq$ by minimizing $\Pi_0$: $\left. \frac{\partial \Pi_0}{\partial (aq)} \right|_{cq, \varepsilon, W_A h^2/D} = 0$. It is important to note also that in curves 1 and 2 of figure 2, normal and shear stresses remain concentrated within a distance equivalent to one wavelength of the oscillation i.e. $\approx 5q^{-1}$, for curves 3 and 4, this distance spans to half the width of the contact area i.e. $c$. Therefore, for these profiles, the relevant characteristic length along $x$ is not $q^{-1}$ but $c$ which motivates us to redefine the confinement parameter as $h/c$.

**Geometric Perturbation Analysis:**

We now discuss the situation when the contact line becomes undulatory. The effect of confinement on the contact line instability is evident in the video-micrographs 1(b)-(e), which show the typical images of the contact region with increasing contact width $2c$. No undulation can be observed when the distance $2c$ decreases below a critical value; however, they appear along the contact line when the contact width is increased and their amplitude progressively increases with further increase in $2c$. We measure the amplitude $A$ of these waves, after the contact line stops completely and normalize it as $\xi = Aq$. These data summarized in figure 1f show that the scaled amplitude $\xi$ varies inversely with the confinement parameter $h/c$. The amplitude, however, does not increase from zero but from a finite value at the critical confinement, $h/c = 0.14$. No undulations could be observed beyond this limit.

In order to estimate the threshold confinement at which non-trivial solutions of excess quantities $(T_i, i \neq 0)$ are energetically favorable, we solve the following equations obtained by matching the coefficients for $\varepsilon^i, i = 2, 4$ in the left and the right hand side of equation 7:

$$\text{(a)}\ \varepsilon^2 : \frac{\partial P_1}{\partial X} = \frac{\partial^2 U_1}{\partial Y^2} + \frac{\partial^2 U_1}{\partial Z^2}, \qquad \frac{\partial P_1}{\partial Y} = 0, \qquad \frac{\partial P_1}{\partial Z} = 0$$

$$\text{(b)}\ \varepsilon^4 : \frac{\partial P_2}{\partial X} = \frac{\partial^2 U_1}{\partial X^2} + \frac{\partial^2 U_2}{\partial Y^2} + \frac{\partial^2 U_2}{\partial Z^2}, \quad \frac{\partial P_2}{\partial Y} = \frac{\partial^2 V_1}{\partial Y^2} + \frac{\partial^2 V_1}{\partial Z^2}, \quad \frac{\partial P_2}{\partial Z} = \frac{\partial^2 W_1}{\partial Y^2} + \frac{\partial^2 W_1}{\partial Z^2} \qquad (13)$$

$$\text{(c)}\ \frac{\partial U_1}{\partial X} + \frac{\partial V_1}{\partial Y} + \frac{\partial W_1}{\partial Z} = 0$$

Similarly, the boundary conditions in equation 8 result in,

(a) at $Z = 0$, $\quad \varepsilon^2 : U_1 = V_1 = W_1 = 0, \quad \varepsilon^4 : U_2 = V_2 = W_2 = 0$

(b) at $Z = 1 + \Psi$, $\quad \varepsilon^2 : \frac{\partial U_1}{\partial Z} = \left( \frac{\partial V_1}{\partial Z} + \frac{\partial W_1}{\partial Y} \right) = 0 \quad \varepsilon^4 : \frac{\partial U_2}{\partial Z} + \frac{\partial W_1}{\partial X} = \frac{\partial V_2}{\partial Z} + \frac{\partial W_2}{\partial Y} = 0$

(c) at $Z = 1 + \Psi$ and $X < \mathrm{N}(Y)$, $\quad \varepsilon^{-2} : 0 = \dfrac{\partial^4 \Psi_1}{\partial Y^4}$

$$\varepsilon^0 : \; 0 = 2\frac{\partial^4 \Psi_1}{\partial X^2 \partial Y^2} + \frac{\partial^4 \Psi_2}{\partial Y^4}$$

$$\varepsilon^2 : P_1 = 3\frac{\partial^4 \Psi_1}{\partial X^4} + 2\frac{\partial^4 \Psi_2}{\partial X^2 \partial Y^2}$$

(d) at $Z = 1 + \Psi$ and $\mathrm{N}(Y) < X < aq$, $\quad \varepsilon^2 : \; 0 = \dfrac{\partial^4 \Psi_1}{\partial X^4}$ (14)

We further assume that the excess displacements and pressure vary sinusoidally along $Y$: $T_i = \overline{T_i} \sin(KY)$; $T = U$, $W$ and $P$ and $V_i = \overline{V_i} \cos(KY)$: $i = 1, 2, \ldots$; where $K = 2\pi / (\lambda / h)$ is the dimensionless wave number of the perturbed waves and $\lambda$ is the wavelength. Notice that the long-scale approximations used for estimating the non-perturbative (i.e. when the contact line is not undulatory) solutions are not relevant here for obtaining the expressions for the perturbed components, because over here, the simplification is effected by mapping the coefficients of $\varepsilon^i, i = 2, 4$ in either sides of the equations 7. Using these new definitions for the excess quantities in equation 13, we obtain the following equations:

(a) $\varepsilon^2 : \; \dfrac{d\overline{P_1}}{dX} = -K^2 \overline{U_1} + \dfrac{d^2 \overline{U_1}}{dZ^2}$, $\quad \overline{P_1} = 0$, $\quad \dfrac{d\overline{P_1}}{dZ} = 0$

(b) $\varepsilon^4 : \; \dfrac{d\overline{P_2}}{dX} = \dfrac{d^2 \overline{U_1}}{dX^2} - K^2 \overline{U_2} + \dfrac{d^2 \overline{U_2}}{dZ^2}$, $\quad K\overline{P_2} = -K^2 \overline{V_1} + \dfrac{d^2 \overline{V_1}}{dZ^2}$,

$$\frac{d\overline{P_2}}{dZ} = -K^2 \overline{W_1} + \frac{d^2 \overline{W_1}}{dZ^2} \qquad (15)$$

(c) $\dfrac{d\overline{U_1}}{dX} - K\overline{V_1} + \dfrac{d\overline{W_1}}{dZ} = 0$

which are solved using the following boundary conditions obtained from equation 14a-b:

(a) at $Z = 0$, $\varepsilon^2 : \overline{U_1} = \overline{V_1} = \overline{W_1} = 0$, $\varepsilon^4 : \overline{U_2} = \overline{V_2} = \overline{W_2} = 0$

(b) at $Z = 1 + \Psi$, $\varepsilon^2 : \frac{d\overline{U_1}}{dZ} = \left(\frac{d\overline{V_1}}{dZ} + K\overline{W_1}\right) = 0$, $\varepsilon^4 : \frac{d\overline{U_2}}{dZ} + \frac{d\overline{W_1}}{dX} = \frac{d\overline{V_2}}{dZ} + K\overline{W_2} = 0$ (16)

Equation 15(a) suggests that the $\varepsilon^2$ order of the excess pressure $P_1$ is zero, whereas that for the $\varepsilon^4$ order of the pressure $P_2$ is finite, implying that the film undergoes undulations at the surface under a very small excess pressure. This excess pressure, however small, varies along $Y$, signifying that it depends upon the gap between the plate and film [13, 25]. Nevertheless, this excess pressure applies only in the immediate vicinity (<0.1 μm) of the contact between the film and plate as the gap between the two increases rather sharply as can be seen in the atomic force microscopy (AFM) images presented in figure 1(g); it does not contribute any significantly to the overall energy of the final state of the system. Therefore, in this calculation we find only the $\varepsilon^2$ order of the excess displacements and pressure. These are obtained by solving equations 15 subjected to the boundary conditions 16a-b relevant for the film; as for the plate, its excess vertical displacement is such that it remains in contact with the film only through one half of the wave. Hence, a simple way of expressing the excess displacement of the plate is

$$\Psi_1 = \overline{W_1}(X, Z = 1 + \Psi) + \Psi_{11}(X)(1 - \sin(KY))$$

and $\Psi_2 = \overline{W_2}(X, Z = 1 + \Psi) + \Psi_{22}(X)(1 - \sin(KY))$

in which $\Psi_{11}$ and $\Psi_{22}$ are the amplitudes. The above definition implies that the plate remains in contact with the film at $KY = (2n+1)\pi/2$, $n = 0, 2, 4, \ldots$, at all other values of $KY$ it remains out of contact. Notice that this definition ensures only a line contact between the plate and the film, this has to do with the fact that here we consider only one component of the Fourier spectrum. A more accurate approach would be to consider all the components which would ensure contact whole through one half of the wave as was done rather elegantly by Sarkar et al [25] while modeling the debonding of a rigid plate from a confined elastic layer bonded to a rigid substrate. However, at the onset of appearance of the perturbation, it is sufficient to consider only a single wave vector of the whole spectrum in order to obtain the condition of occurrence of the instability. We assume further that $\Psi_{11} << \overline{W_1}$ and $\Psi_{22} << \overline{W_2}$, so that we can write directly $\Psi_1 = \overline{W}_1(X, Z = 1 + \Psi)$ and $\Psi_2 = \overline{W}_2(X, Z = 1 + \Psi)$ without sacrificing much accuracy. In essence, this assumption implies that the plate bends only along the $X$ axis, remaining uniform along $Y$, as we observed in our experiments. Then the boundary conditions 14c and 14d simplify to the following equations for the excess displacement of the plate:

$$\text{at } Z = 1 + \Psi, \qquad X < \mathrm{N}(Y) \quad 0 = \frac{\partial^4 \Psi_1}{\partial X^4} \quad \text{at } \quad \mathrm{N}(Y) < X < aq \quad 0 = \frac{\partial^4 \Psi_1}{\partial X^4} \tag{17}$$

In essence, the above condition is applicable at any $X$, then it is convenient to apply the following boundary conditions at $X = 0$, rather than at $X = \mathrm{N}(Y)$ :

The excess displacement $\Psi_1$ and its slope are continuous at $X = 0$ so that $\Psi_1|_{X=0-} = \Psi_1|_{X=0+} = C_0 \Phi(K), \left.\frac{\partial \Psi_1}{\partial X}\right|_{X=0-} = \left.\frac{\partial \Psi_1}{\partial X}\right|_{X=0+}$ and $\left.\frac{\partial^2 \Psi_1}{\partial X^2}\right|_{X=0-} = \left.\frac{\partial^2 \Psi_1}{\partial X^2}\right|_{X=0+}$. Here, $C_0 \Phi(K)$ is the excess stretching of the film at $X = 0$ (Reference 13), $C_0$ is a constant and the function $\Phi(K)$ falls out naturally from the solution of the excess quantities. At $X = -\xi$, where, $\xi = Aq$ is the dimensionless amplitude of the waves, the excess displacement of the plate, its slope vanishes, which result in the b.c.

$\Psi_1\big|_{X=-\xi} = \left.\frac{\partial \Psi_1}{\partial X}\right|_{X=-\xi} = 0$. Similarly, at $X = aq$ excess displacement and curvature of plate is zero, which results in b.c. $\Psi_1\big|_{X=aq} = \left.\frac{\partial^2 \Psi_1}{\partial X^2}\right|_{X=aq} = 0$. Equation 15 and the boundary conditions discussed thus far, dictate the analytical expressions of the excess quantities which are same as those obtained for the experiment with single spacer and have been solved in reference 16.

**Excess energy:**

After obtaining the expressions for the base and the excess displacements [16], we proceed to estimate the excess energy of the system $\Pi = \Pi_{Total} - \Pi_0$. Using $\mu/\varepsilon^3$ as the characteristic energy and considering only the leading order terms ($\varepsilon^4$ and $\varepsilon^6$) we obtain the expression for the excess energy as,

$$\begin{aligned}\Pi = {} & \frac{\varepsilon^6}{4}\int_{-\xi}^{0}\int_{0}^{2\pi/K}\int_{0}^{1}\left(\frac{\partial V_1}{\partial Z} + \frac{\partial W_1}{\partial Y}\right)^2 dZ\, dY\, dX \; + \\ & \frac{3\pi}{K}\int_{-\xi}^{aq}\left(2\varepsilon^2 \frac{\partial^2 \Psi_0}{\partial X^2}\frac{\partial^2 \Psi_1}{\partial X^2} + \varepsilon^4\left(\frac{\partial^2 \Psi_1}{\partial X^2}\right)^2\right) dX + \frac{6\xi}{K}\frac{W_A h^2}{D\varepsilon^4}\end{aligned} \tag{18}$$

The first term in the right side corresponds to the excess elastic energy of the film, estimated only within a distance $-\xi \le X \le 0$ at which the surface of the film gets undulatory. The second term corresponds to the bending energy of the plate which needs to be estimated within a distance $-\xi \le X \le aq$. The third term represents the excess interfacial energy associated with the planar area of the fingers. We have neglected however the interfacial energy associated with their side walls.

After substituting the expressions for the base component of displacement $\Psi_0$ and the excess displacements $V_1, W_1$ and $\Psi_1$, we finally obtain

$$\Pi = \left(\varepsilon^6 f_2(\xi, cq, K) + \varepsilon^4 f_1(\xi, cq, K)\right) C_0^2 + \varepsilon^2 f_3(\xi, cq, K)\overline{\Delta} C_0 + \frac{6\xi}{K}\frac{W_A h^2}{D\varepsilon^4} \quad (19)$$

Here, the expressions for $f_1(\xi, cq, K), f_2(\xi, cq, K)$ and $f_3(\xi, cq, K)$ are obtained using Mathematica. Furthermore, hypothesizing that $\Pi$ minimizes when $\partial\Pi/\partial C_0 = 0$, we obtain an expression for $C_0$ which when substituted in equation 19, yields

$$\Pi = -\frac{\overline{\Delta}^2}{4}\frac{f_3^2}{f_1 + \varepsilon^2 f_2} + \frac{6\xi}{K}\frac{W_A h^2}{D\varepsilon^4} \quad (20)$$

Thus the excess energy $\Pi$ is obtained as a function of three types of parameters: the length-scales of perturbation (i.e. wavelength $\lambda / h$ and amplitude $\xi = Aq$), geometric length-scales of the experiment (i.e. $h / c$ and $\varepsilon$), and dimensionless work of adhesion (i.e. $W_A h^2 / D$). Figure 3a depicts the dimensionless excess energy as a function of dimensionless wave number $K = 2\pi h / \lambda$ and amplitude $\xi$ while $h / c$, $\varepsilon$ and $W_A h^2 / D$ are kept constant. $\Pi$ goes through a minimum ($\Pi_{\min}$) which remains positive as long as $h / c$ is greater than a critical value. However, when $h / c < (h / c)_c$, we obtain a negative minimum for $\Pi_{\min}$, implying that undulations with finite amplitude are energetically favorable at and beyond this confinement. The corresponding values of $\lambda / h$, plotted against $h / c$ in figure 3b for a representative set of values of $\varepsilon = 0.19 - 0.35$ and $\left(W_A h^2 / D\right)^{1/4} = 0.01 - 0.015$ show that undulations appear only when the film is sufficiently confined i.e. $h / c < 0.13$ as is observed in experiments (figure 1f) with variety of films of different thickness and modulus and cover plates of different rigidity.

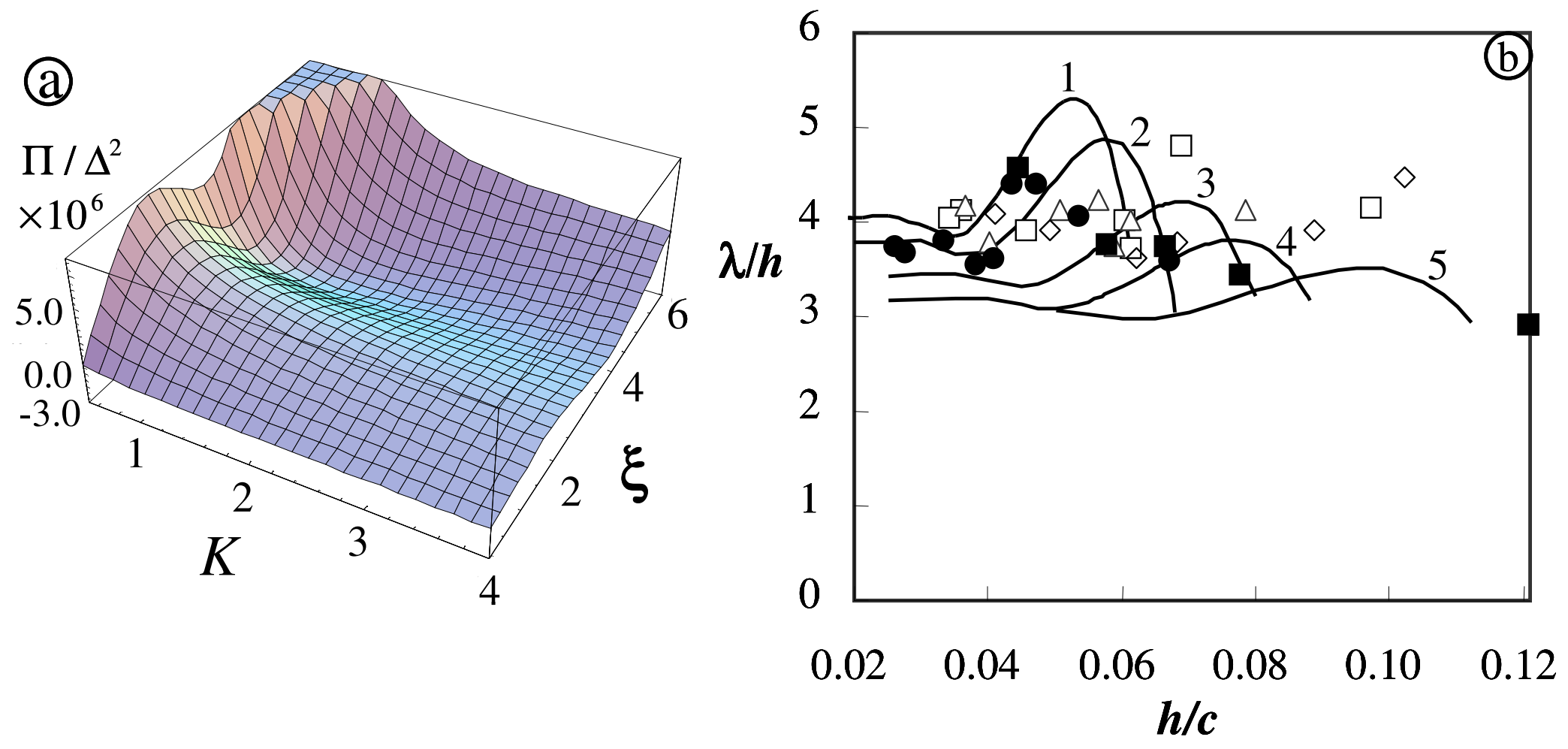

**Fig 3. (a) (Color online) (a) A typical plot of dimensionless excess energy $\Pi$ as a function of wave-number $K = 2\pi h/\lambda$ and amplitude $\xi = Aq$ of fingers for dimensionless parameters: $\varepsilon = 0.29$, $h/c = 0.084$ and $\left(W_A h^2 / D\right)^{1/4} = 0.008$. Instability ensues when $\Pi$ is negative and minimum. (b) These plots are obtained for variety of $\varepsilon$ and $h/c$ at a given $\left(W_A h^2 / D\right)^{1/4}$ and extracting from them the data of $\lambda/h$ at which $\Pi = \Pi_{\min}$ and it is negative. The solid lines (1 to 5) represent these data of $\lambda/h$ plotted against $h/c$ for variety of values for $\varepsilon = 0.192, \left(W_A h^2 / D\right)^{1/4} = 0.01$, $(0.21, 0.011)$, $(0.252, 0.0122)$, $(0.282, 0.0147)$ and $(0.35, 0.01)$ respectively. The symbols □, ◊, Δ, ■ and ● represent the data obtained from experiments for $\varepsilon = 0.189$ and $\left(W_A h^2 / D\right)^{1/4} = 0.01$, $(0.186, 0.0114)$, $(0.166, 0.009)$, $(0.252, 0.0122)$ and $(0.133, 0.0082)$ respectively.**

The wavelength of instability varies non-monotonically with the contact width which qualitatively captures the experimental observation that $\lambda/h$ does not show monotonic increase or decrease with $h/c$ as shown in figure 3b. Furthermore, in the limit $h/c \to 0$, i.e. for large enough contact width, $\lambda/h$ becomes independent of $h/c$ as the experiment then converges to the single spacer geometry. The theory captures also our experimental observation that the amplitude does not increase from zero, but from a finite value, so that only perturbations with finite amplitude grow while others decay. However amplitude is somewhat overestimated which could be due to the assumption of sinusoidal variation of the excess quantities along the $y$ axis.

**Prestretched films:**

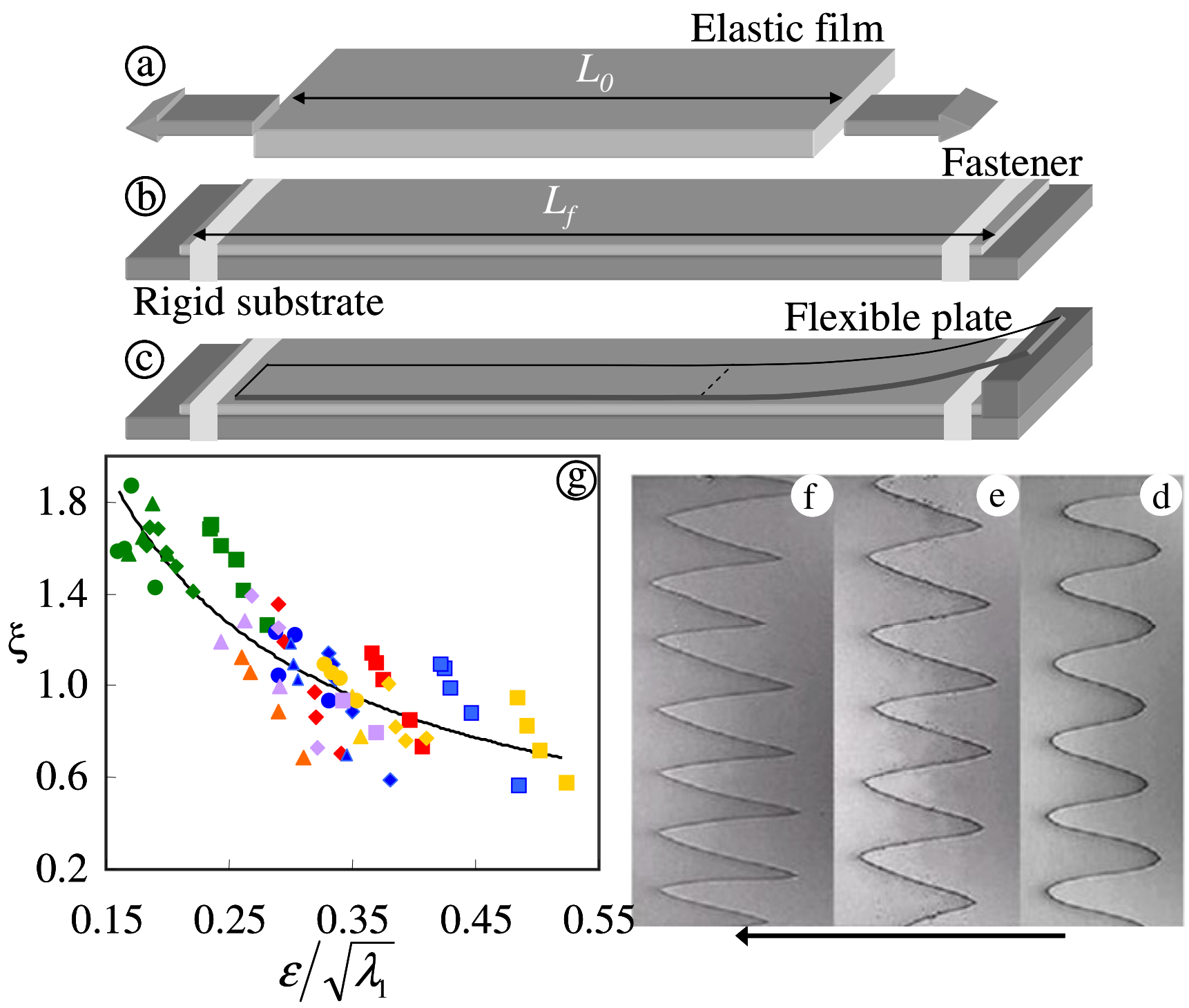

**Fig 4. (Color online) Pattern formation with pre-stretched films. A uniaxially stretched thin film of sylgard 184 is clamped to a rigid substrate using an adhesive tape from which a flexible plate is peeled off using a spacer. (d)-(f) Video-micrographs of patterns obtained with a plate of rigidity $D = 0.02$ Nm and a film of modulus $\mu = 1.0$ MPa and thickness $h = 210\,\mu\text{m}$ stretched to $\lambda_1 = 1.4, 1.67$ and $1.87$ respectively. The arrow shows the direction in which the crack opens. (g) The amplitude data from experiments with different $h$, $D$ and $\lambda_1$ are normalized as $\xi = Aq$ and are plotted against the modified confinement parameter $\varepsilon / \sqrt{\lambda_1}$. The solid curve is a guide to eye and captures the overall variation of $\xi$ with $\bar{\varepsilon}$.**

While so far we discussed about elastic films free of any pre-stress, here we will describe our experiments with the ones which are subjected to uniaxial pre-stretching. Figure 4(a)-(c) depict the schematic in which an elastic film is first stretched along $x$ and is then placed on a rigid substrate on which it remains strongly adhered. A flexible plate is then peeled off the film. If $\lambda_1$ is the extension ratio of the film along $x$, i.e. $\lambda_1 = L_f / L_0$ ($\lambda_1 > 1$), then due to incompressibility, extension ratio along the other two directions is $\lambda_2 = 1/\sqrt{\lambda_1}$. Video-micrographs in figure 4(d)-(f) depict the typical examples of instability patterns which appear along the contact line. Here ”V” shaped fingers appear unlike the ”U” shaped ones as seen in previous experiments. While, these patterns were detected earlier by Lake et al [26] during low angle peeling of a adhesive and also by others [27] on a different context, here we study them systematically by subjecting the polymer films of different thicknesses to different extension ratios and by peeling off flexible plates of varying rigidity.

For example, in figure 4d-f, a film with $\mu = 1.0$ Mpa and $h = 210\,\mu$m is progressively stretched to $\lambda_2 = 0.847, 0.774$ and $0.732$ respectively and a plate of $D = 0.02$ Nm is peeled from it. The resultant fingers have round tips when $\lambda_2$ is closed to 1.0, however, with increasing extension of the film sharpness of the tip increases. Furthermore, the increase of the amplitude of the fingers with decreasing $\lambda_2$ implies that increase in pre-extension leads to longer fingers. Similar to our earlier observations, amplitude increases also with the rigidity of the contacting plate and decreases as the thickness of the film is increased.

These features can be rationalized by considering the equilibrium of incremental stresses [28] in an incompressible elastic layer as follows (please see **Appendix A** for derivation):

$$\begin{aligned}
&\frac{1}{\mu}\frac{\partial s}{\partial x}+\frac{1}{3}\left(\frac{2}{\lambda_1}+\lambda_1^2\right)\frac{\partial^2 u}{\partial x^2}+\frac{1}{\lambda_1}\frac{\partial^2 u}{\partial z^2}=0\\
&\frac{1}{\mu}\frac{\partial s}{\partial z}+\lambda_1^2\frac{\partial^2 w}{\partial x^2}+\frac{1}{3}\left(\frac{1}{\lambda_1}+2\lambda_1^2\right)\frac{\partial^2 w}{\partial z^2}=0
\end{aligned} \tag{21}$$

Here $u$ and $w$ are the incremental deformations along $x$ and $z$, whereas $s$ is the incremental average stress. While the effect of pre-extension is intrinsically accounted for in equation 21 it is not amenable to analytical solutions except at the limiting situations: $\lambda_1 \to 1$ and $\lambda_1 >> 1$. When $\lambda_1 \to 1$, long-scale approximations can be used to simplify equation 21 as $\frac{1}{\mu}\frac{\partial s}{\partial x}+\frac{1}{\lambda_1}\frac{\partial^2 u}{\partial z^2}=\frac{\partial s}{\partial z}=\frac{\partial u}{\partial x}+\frac{\partial w}{\partial z}=0$, which is integrated using the conditions: perfect bonding at $z=0$; frictionless contact and continuity of normal stresses at $z=h/\sqrt{\lambda_1}$ (due to stretching, thickness decreases to $h/\sqrt{\lambda_1}$ ). This procedure yields a relation:

$$\psi_0=\frac{Dh^3}{3\mu\sqrt{\lambda_1}}\frac{\partial^6\psi_0}{\partial x^6} \text{ at } x<0 \tag{22}$$

from which we obtain the characteristic lengths, along $x$:

$\left(\frac{Dh^3}{3\mu\sqrt{\lambda_1}}\right)^{1/6}\approx q^{-1}$ and along $z$: $h/\sqrt{\lambda_1}$. The confinement parameter: $\bar{\varepsilon}=\varepsilon/\sqrt{\lambda_1}$

then implies that pre-stretching enhances the confinement of the film so that undulations can appear even for thicker films when sufficiently pre-stretched. The video micrographs in figure 5a-d illustrate this situation, in which a plate of rigidity $D=0.02$ Nm is lifted from an elastic film ( $h=645\,\mu$m and $\mu=1.0$ Mpa). The contact line remains straight till $\lambda_1 < 1.38$ beyond which it turns undulatory. The critical value of the confinement parameter is then obtained as $\bar{\varepsilon} \approx 0.52$ which is rather large. However, in the limit of $\lambda_1 >> 1$, simplification of equation 21 with the assumption of constant $s$ along $z$ leads to

$$\psi_0 + \frac{3Dh}{\mu} \frac{\sqrt{\lambda_1}}{\left(\lambda_1^3 + 2\right)} \frac{\partial^4 \psi_0}{\partial x^4} = 0 \text{ at } x < 0 \tag{23}$$

which results in a different definition: $\bar{\varepsilon} = \left( \frac{\mu h^3}{3D} \frac{\lambda_1^3 + 2}{\sqrt{\lambda_1}} \right)^{1/4}$ and a critical value of confinement parameter $\bar{\varepsilon} = 0.36$ which is similar to our earlier observations with un-stretched films ($\varepsilon = 0.35$) [14]. Furthermore the scaled amplitude $\xi = Aq$ from variety of experiments ($h = 245 - 760$ μm, μ = 1.0 Mpa, $D = 0.02 - 0.42$ Nm and $\lambda_1 = 1.45 - 1.93$) increases with decrease in $\bar{\varepsilon}$ (figure 4g) as observed earlier. The scaled wavelength, however, does not remain constant, it increases from 2.25 to 4.25.

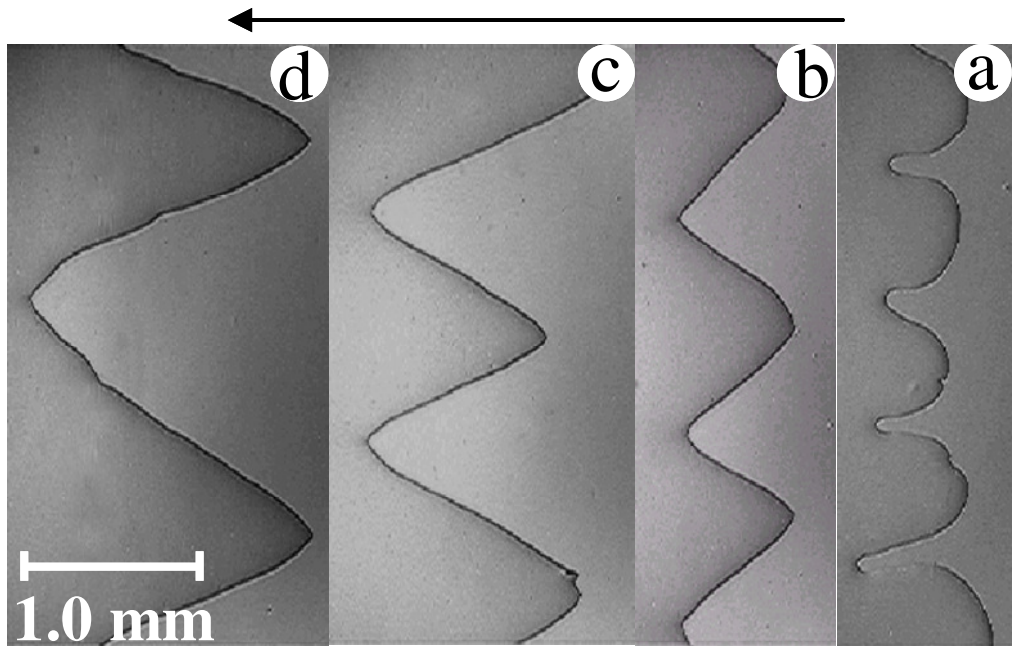

**Fig 5. Video micrographs of patterns observed with Sylgard-184 elastomeric film of thickness *h* = 645 μm and shear modulus μ = 1.0 Mpa and stretched to $\lambda_1 = 1.45$. Micrographs (a)-(d) are obtained with glass plates of rigidity,** $D = 0.02,\ 0.09,\ 0.154,\ 0.21$ **Nm respectively.**

**Summary:**

In this report we have examined the effect of confinement of a thin elastic film via two different experiments in which we have varied systematically the confinement by controlling the width of contact between the film and a flexible plate and by subjecting the film to pre-extension. We have characterized the confinement of the film by the ratio of two characteristic lengthscales: thickness of the film and the distance from the contact line within which the stresses remain concentrated. Whereas for experiments with semi-infinite contact width (i.e. with single spacer), the confinement parameter is obtained as $\varepsilon = hq$; it is obtained as $h/c$ when contact width is finite (as in figure 1) and $hq/\sqrt{\lambda_1}$ for pre-stretched films (figure 4). We show that in this variety of situations, the instability in the form of finger like patterns appears at the contact line when a threshold confinement is attained and the amplitude increases with further increase in confinement. Furthermore, in all these experiments, the amplitude increases not from zero but from a finite value. These features are all captured through our analysis in which we estimate the excess energy of the system associated with the appearance of these undulations. We show that the instability should indeed ensue with finite amplitude as the minima of the excess energy turns negative at a critical value of the confinement parameter. Our analysis however depicts a simplified picture of the very complex displacement and stress fields in the vicinity of the contact line; because, over here, we consider only a single Fourier mode in describing the transverse modulations of the displacements which fails to account for the detailed 3d morphology of the interface. A full simulation of the problem coupled with a Fourier series expansion of the undulation should capture the detailed morphology of the instability patterns beyond critical confinement.

**Acknowledgement**: Part of this work constituted the PhD thesis of A. Ghatak at Lehigh University. We acknowledge the Office of Naval Research and the Pennsylvania Infrastructure Technology Alliance (PITA) for financial supports.

*Electronic address: aghatak@iitk.ac.in

**Appendix A:**

The detail analysis of incremental deformations can be found in reference [25] from where we obtain the stress equilibrium relations in terms of the incremental stresses in an elastic material which is already subjected to pre-stresses. Let us say that an elastic body is subjected to the initial stress field:

$$\begin{matrix} S_{11} & S_{12} & S_{13} \\ S_{12} & S_{22} & S_{23} \\ S_{13} & S_{23} & S_{33} \end{matrix}$$

where directions $1, 2$ and $3$ correspond to $x$, $y$ and $z$ respectively. If the body is now deformed, a new stress field develops in the body, which when expressed with respect to the axes that rotate with the medium can be written as,

$$\begin{matrix} S_{11}+s_{11} & S_{12}+s_{12} & S_{13}+s_{12} \\ S_{12}+s_{12} & S_{22}+s_{22} & S_{23}+s_{23} \\ S_{13}+s_{13} & S_{23}+s_{23} & S_{33}+s_{33} \end{matrix}$$

Here $s_{11}$, $s_{22}$ and $s_{12}$ etc. are the incremental stresses. In our problem an incompressible elastic film is pre-stretched uniaxially, so that $\lambda_1$ is the extension ratio, $\lambda_1 = L_f / L_0$ ($\lambda_1 > 1$), then due to incompressibility, extension ratio along the other two directions are $\lambda_2 = \lambda_3 = 1/\sqrt{\lambda_1}$. Using the neo-Hookean model, the initial stress $S_{11}$ in the material along direction $x$ can be written as:

$$S_{11} = \mu\left(\lambda_1^2 - \frac{1}{\lambda_1}\right) \tag{A1}$$

The other components of the initial stresses are zero, i.e. $S_{12} = S_{13} = S_{23} = S_{33} = S_{22} = 0$.

In the absence of body forces, the simplified form of the stress equilibrium relations (see equation 7.49 in page 52 of reference 25) are:

$$
\begin{aligned}
&\frac{\partial s_{11}}{\partial x}+\frac{\partial s_{12}}{\partial y}+\frac{\partial s_{13}}{\partial z}+S_{11}\left(\frac{\partial \omega_z}{\partial y}-\frac{\partial \omega_y}{\partial z}\right)=0\\
&\frac{\partial s_{12}}{\partial x}+\frac{\partial s_{22}}{\partial y}+\frac{\partial s_{23}}{\partial z}+S_{11}\frac{\partial \omega_z}{\partial x}=0\\
&\frac{\partial s_{13}}{\partial x}+\frac{\partial s_{23}}{\partial y}+\frac{\partial s_{33}}{\partial z}-S_{11}\frac{\partial \omega_y}{\partial x}=0
\end{aligned}
\tag{A2}
$$

and the incompressibility relation is $e_{xx}+e_{yy}+e_{zz}=0$.

The incremental stress components are expressed in terms of strains as (see equation 8.46 in page 104 of reference 25),

$$
\begin{aligned}
&s_{11}-s_{22}=\frac{2\mu}{\lambda_1}\left(e_{yy}-e_{zz}\right),\quad s_{22}-s_{33}=2\mu\left(\frac{1}{\lambda_1}e_{zz}-\lambda_1^2 e_{xx}\right)\\
&s=\quad \frac{1}{3}\left(s_{11}+s_{22}+s_{33}\right)\\
&s_{12}=\mu\left(\lambda_1^2+\frac{1}{\lambda_1}\right)e_{xy},\quad s_{23}=\frac{2\mu}{\lambda_1}e_{yz},\quad s_{31}=\mu\left(\lambda_1^2+\frac{1}{\lambda_1}\right)e_{xz}
\end{aligned}
\tag{A3}
$$

and, the incremental strain and rotational components are

$$
\begin{aligned}
&e_{xx}=\frac{\partial u}{\partial x},\qquad e_{xy}=\frac{1}{2}\left(\frac{\partial u}{\partial y}+\frac{\partial v}{\partial x}\right),\quad \omega_z=\frac{1}{2}\left(\frac{\partial v}{\partial x}-\frac{\partial u}{\partial y}\right)\\
&e_{yy}=\frac{\partial v}{\partial y},\qquad e_{zx}=\frac{1}{2}\left(\frac{\partial w}{\partial x}+\frac{\partial u}{\partial z}\right),\quad \omega_y=\frac{1}{2}\left(\frac{\partial u}{\partial z}-\frac{\partial w}{\partial x}\right)\\
&e_{zz}=\frac{\partial w}{\partial z},\qquad e_{yz}=\frac{1}{2}\left(\frac{\partial v}{\partial z}+\frac{\partial w}{\partial y}\right),\quad \omega_x=\frac{1}{2}\left(\frac{\partial w}{\partial y}-\frac{\partial v}{\partial z}\right)
\end{aligned}
\tag{A4}
$$

Using the definitions in equation (A1), (A3) and (A4) we can rewrite equation (A2) as,

$$
\begin{aligned}
&\frac{\partial s}{\partial x}+\frac{\mu}{3}\left(\lambda_1^2+\frac{2}{\lambda_1}\right)\frac{\partial^2 u}{\partial x^2}+\frac{\mu}{\lambda_1}\left(\frac{\partial^2 u}{\partial y^2}+\frac{\partial^2 u}{\partial z^2}\right)=0\\
&\frac{\partial s}{\partial y}+\mu\lambda_1^2\frac{\partial^2 v}{\partial x^2}+\frac{\mu}{\lambda_1}\left(\frac{\partial^2 v}{\partial y^2}+\frac{\partial^2 v}{\partial z^2}\right)+\frac{2\mu}{3}\left(-\lambda_1^2+\frac{1}{\lambda_1}\right)\frac{\partial^2 u}{\partial x\partial y}=0\\
&\frac{\partial s}{\partial z}+\mu\lambda_1^2\frac{\partial^2 w}{\partial x^2}+\frac{\mu}{\lambda_1}\left(\frac{\partial^2 w}{\partial y^2}+\frac{\partial^2 w}{\partial z^2}\right)+\frac{2\mu}{3}\left(-\lambda_1^2+\frac{1}{\lambda_1}\right)\frac{\partial^2 u}{\partial x\partial z}=0
\end{aligned}
\tag{A5}
$$

In order to obtain the base solutions, we can simplify equation A5 using long scale approximation,

$$
\begin{aligned}
&\frac{1}{\mu}\frac{\partial s}{\partial x}+\frac{1}{3}\left(\lambda_1^2+\frac{2}{\lambda_1}\right)\frac{\partial^2 u}{\partial x^2}+\frac{1}{\lambda_1}\frac{\partial^2 u}{\partial z^2}=0\\
&\frac{1}{\mu}\frac{\partial s}{\partial z}+\lambda_1^2\frac{\partial^2 w}{\partial x^2}+\frac{1}{3}\left(2\lambda_1^2+\frac{1}{\lambda_1}\right)\frac{\partial^2 w}{\partial z^2}=0
\end{aligned}
\tag{A6}
$$

Equation A6 is used for analyzing the problem with pre-stretched films.